\documentclass[preprint, superscriptaddress, showpacs,preprintnumbers,amsmath,amssymb]{revtex4}
\usepackage{graphicx}

\begin{document}

\thispagestyle{empty}

\title{Constraints on axion and corrections to Newtonian
gravity from the Casimir effect\footnote{Based on two 
 talks presented by V.~M.~Mostepanenko
and G.~L.~Klimchitskaya at the 15th Russian Gravitational Conference 
--- International Conference on Gravitation, Cosmology and 
Astrophysics (RUSGRAV-15). June 30--July 5, 2014. Kazan, Russia.}}

\author{
G.~L.~Klimchitskaya\footnote{E-mail: g.klimchitskaya@gmail.com}
}
\affiliation{Central Astronomical Observatory
at Pulkovo of the Russian Academy of Sciences,
St.Petersburg, 196140, Russia}
\affiliation{Institute of Physics, Nanotechnology and
Telecommunications, St.Petersburg State
Polytechnical University, St.Petersburg, 195251, Russia}
\author{V.~M.~Mostepanenko\footnote{E-mail: vmostepa@gmail.com}
}
\affiliation{Central Astronomical Observatory
at Pulkovo of the Russian Academy of Sciences,
St.Petersburg, 196140, Russia}
\affiliation{Institute of Physics, Nanotechnology and
Telecommunications, St.Petersburg State
Polytechnical University, St.Petersburg, 195251, Russia}

\begin{abstract}
Axion is a light pseudoscalar particle of much interest for physics of elementary
particles and for astrophysics.
We review the recently obtained  constraints on axion to nucleon coupling constants
following from different experiments
on measuring the Casimir interaction.
These constraints are compared with those following from other laboratory
experiments within the wide range of masses of axion-like particles from
$10^{-10}$ to 20\,eV.
We also collect the most strong constraints on the Yukawa-type and power-type
corrections to the Newton law of gravitation which follow from measurements of
the Casimir interaction, E\"{o}tvos- and Cavendish-type experiments.
The possibility to obtain stronger constraints on an axion from the Casimir effect
is proposed.
\end{abstract}

\maketitle
\section{Introduction}

The light pseudoscalar particle named {\it axion} is an important element of the
Standard Model and its generalizations. Axion arises \cite{1,2} due to breaking
of the Peccei-Quinn symmetry which was introduced \cite{3} in quantum
chromodynamics (QCD) in order to avoid strong {\it CP} violation and large
electric dipole moment of a neutron (numerous experiments exclude both these
effects  to a high level of precision).
What is more, axions provide an elegant solution for the problem of dark matter in
astrophysics and cosmology \cite{4,5}. This is the reason why a lot of experiments for
searching axions has been performed in different countries \cite{6}.
Specifically, strong constraints on the coupling constants of an axion and other axion-like
particles with photons, electrons and nucleons were obtained from astrophysical
observations. Up to the present, however, there is the so-called {\it window} in the
values of an axion mass, where these constraints are either missing or not sufficiently
strong.

There are also massless and light scalar particles predicted in many extensions of the
Standard Model \cite{7}. Exchange of such particles between atoms of two macrobodies
leads to corrections to the Newton law of gravitation at separations below a micrometer.
By coincidence, at so small separations Newton's gravitational law is not verified experimentally
with sufficient precision. Within a submicrometer interaction range experiment does not
exclude corrections which exceed the Newton gravitational force by many orders of
magnitude \cite{8}. Similar corrections are predicted in extra-dimensional models with a
low-energy compactification scale \cite{9,10}. Many experiments of E\"{o}tvos- and Cavendish-type have
been performed during the last few years searching for possible corrections to the Newton
law of gravitation \cite{11}.

Recently it was found \cite{12,13,14,15} that strong model-independent constraints on the coupling
constants of axions with nucleons follow from measurements of the Casimir-Polder and Casimir force.
Some of these constraints overlap with an axion window and, thus, are complementary to
astrophysical limits. As to corrections to Newton's law of gravitation, measurements of the van der Waals
and Casimir forces have long been used to constrain their parameters \cite{16,17}.
New, more precise measurements of the Casimir force allowed significant strengthening of previously
obtained constraints on non-Newtonian gravity over the region of separations below $1\,\mu$m
\cite{18,19,20,21,22}.

In this paper, we review constraints on the coupling constants of an axion to a proton and a neutron,
and corrections to Newton's law of gravitation which follow from the most precise measurements of the
Casimir interaction \cite{23,24}. We compare the obtained constraints on an axion with the alternative
constraints following from some other laboratory experiments. The constraints on the coupling constants
of an axion and on non-Newtonian gravity, following from measurements of the Casimir interaction, are
mutually compared and some conclusions inherent to both of them are obtained.

The paper is organized as follows. In  Section 2 we consider the types of effective potentials which arise
due to one- and two-axion exchange. These are compared with the effective potentials originating from
the exchange of massless and massive scalar particles. Section 3 is devoted to the constraints on
axion-nucleon coupling constants which follow from measurements of the Casimir-Polder force acting
between the condensate of ${}^{87}$Rb atoms and a glass silica plate. In Section 4 the constraints on
axion to nucleon coupling constants are presented obtained from measurements of the gradient of the
Casimir force between a microsphere and a plate coated with a nonmagnetic metal Au or a magnetic
metal Ni. These experiments were performed by means of a dynamic atomic force microscope (AFM).
Section 5 contains similar constraints obtained from measurements of the gradient of the Casimir force
between Au-coated surfaces of a sphere and a plate using a micromachined oscillator.
In Section 6 the constraints on the coupling constants of an axion are provided which follow from
measurements of the Casimir force between corrugated surfaces.  In Section 7 we compare the
constraints on an axion found from measurements of the Casimir interaction with those obtained from
some other laboratory experiments. Section 8 is devoted to the constraints on non-Newtonian gravity
derived from the Casimir effect. In Section 9 the reader will find our conclusions and discussion.

Throughout the paper we use units in which $\hbar=c=1$.

\section{Types of effective potentials}

Below we consider effective potentials arising from the interaction of nucleons (protons and neutrons) with
an axion and other axion-like particles predicted in different variants of the Grand Unification Theories.
Axions also interact with electrons and photons. These interactions are, however, much weaker than
axion-nucleon interaction \cite{25} and for our purposes can be neglected. In any case, their account would
lead to only a minor strengthening of the constraints on axion-nucleon coupling constants obtained from the
force measurements between macroscopic bodies.

We assume that the interaction of axion-like particles $a$ with nucleons $\psi$ is described by the
Lagrangian \cite{4}
\begin{equation}
{\cal L}=-i g_{ak}\bar{\psi}\gamma_5\psi a,
\label{eq1}
\end{equation}
\noindent
where $g_{ak}$ is the coupling constant of an axion to a proton ($k=p$) or to a neutron ($k=n$).
In doing so the pseudoscalar coupling of axions and other axion-like particles to nucleons is assumed
(note that the pseudovector coupling introduced for the
original QCD axions results in the nonrenormalizable
theory \cite{15}).  The exchange of one axion between two nucleons of spins
$\mbox{\boldmath$\sigma$}_{1,2}/2$ situated at the points
$\mbox{\boldmath$r$}_1\neq\mbox{\boldmath$r$}_2$ with coupling (\ref{eq1}) results in the following
effective potential \cite{25,26}
\begin{eqnarray}
&&
V(\mbox{\boldmath$r$}_1-\mbox{\boldmath$r$}_2;\mbox{\boldmath$\sigma$}_{1},\mbox{\boldmath$\sigma$}_{2})
=\frac{g_{ak}g_{al}}{16\pi m_km_l}\left[
\vphantom{\left(
\frac{m_a}{|\mbox{\boldmath$r$}_1-\mbox{\boldmath$r$}_2|^2}
\frac{1}{|\mbox{\boldmath$r$}_1-\mbox{\boldmath$r$}_2|^3}\right)}
(\mbox{\boldmath$\sigma$}_{1}\cdot\mbox{\boldmath$n$})
(\mbox{\boldmath$\sigma$}_{2}\cdot\mbox{\boldmath$n$})\right.
\nonumber\\
&&~~~~~~~~~
\times\left(
\frac{m_a^2}{|\mbox{\boldmath$r$}_1-\mbox{\boldmath$r$}_2|}+
\frac{3m_a}{|\mbox{\boldmath$r$}_1-\mbox{\boldmath$r$}_2|^2}+
\frac{3}{|\mbox{\boldmath$r$}_1-\mbox{\boldmath$r$}_2|^3}\right)
\nonumber\\
&&~~~~
\left.-
(\mbox{\boldmath$\sigma$}_{1}\cdot\mbox{\boldmath$\sigma$}_{2})
\left(
\frac{m_a}{|\mbox{\boldmath$r$}_1-\mbox{\boldmath$r$}_2|^2}+
\frac{1}{|\mbox{\boldmath$r$}_1-\mbox{\boldmath$r$}_2|^3}\right)
\right]\,e^{-m_a|\mbox{\boldmath$r$}_1-\mbox{\boldmath$r$}_2|}.
\label{eq2}
\end{eqnarray}
\noindent
Here, $g_{ak}$ and $g_{al}$ are the axion-proton ($k,l=p$) or 
axion-neutron ($k,l=n$) interaction constants, $m_k,\,m_l$ are the nucleon masses,
$m_a$ is the axion mass, and the unit vector
$\mbox{\boldmath$n$}=
(\mbox{\boldmath$r$}_1-\mbox{\boldmath$r$}_2)
/|\mbox{\boldmath$r$}_1-\mbox{\boldmath$r$}_2|$.

As is seen in (\ref{eq2}), the effective potential depends on the nucleon spins. Because of this,
the resulting interaction between two unpolarized test bodies averages to zero.
Taking into account that already performed experiments on measuring the Casimir interaction
\cite{23,24} deal with unpolarized test bodies, it seems impossible to use them for constraining
the axion to nucleon coupling constants basing on the simplest process of  one-axion exchange.

The situation changes when we consider the process of two-axion exchange between the two
nucleons. In this case the Lagrangian (\ref{eq1}) leads to the following effective
potential \cite{25,27,28}
\begin{equation}
V_{kl}(|\mbox{\boldmath$r$}_1-\mbox{\boldmath$r$}_2|)=-
\frac{g_{ak}^2g_{al}^2}{32\pi^3m_km_l}\,
\frac{m_a}{(\mbox{\boldmath$r$}_1-\mbox{\boldmath$r$}_2)^2}\,
K_1(2m_a|\mbox{\boldmath$r$}_1-\mbox{\boldmath$r$}_2|),
\label{eq3}
\end{equation}
\noindent
where $K_1(z)$ is the modified Bessel function of the second kind. Note that (\ref{eq3}) is
derived under the condition
$|\mbox{\boldmath$r$}_1-\mbox{\boldmath$r$}_2|\gg 1/m_{k,l}$
which is satisfied with a large safety margin in all the experiments considered below.
Equation (\ref{eq3}) does not depend on the nucleon spins. Thus, after the integration over the
volumes of test bodies, it leads to some additional force of the axionic origin which can be
constrained from the measurement results.

Now we address to exchange of massless and light scalar particles between the atoms of two
macroscopic bodies. The exchange of one light scalar particle of mass $M$ between two pointlike
particles with masses $m_1$ and $m_2$ spaced at the points
$\mbox{\boldmath$r$}_1$ and $\mbox{\boldmath$r$}_2$ results in the spin-independent
Yukawa-type effective potential \cite{8}. It is convenient to parametrize this potential as a
correction to Newton's law of gravitation:
\begin{equation}
V(|\mbox{\boldmath$r$}_1-\mbox{\boldmath$r$}_2|)=-
\frac{Gm_1m_2}{|\mbox{\boldmath$r$}_1-\mbox{\boldmath$r$}_2|}\,\left(1+
\alpha e^{-|\mbox{\boldmath$r$}_1-\mbox{\boldmath$r$}_2|/\lambda}\right).
\label{eq4}
\end{equation}
\noindent
Here, $\alpha$ is a dimensionless constant characterizing the strength of Yukawa interaction,
$\lambda=1/M$
 is the Compton wavelength of light scalar particle characterizing the interaction range, and $G$ is the
Newtonian gravitational constant. As was noted in Section 1, the effective potential (\ref{eq4}) arises
also in extradimensional models with a low-energy compactification scale \cite{9,10}.
In this case the quantity $\lambda$ has the meaning of the characteristic size of a multidimensional
compact manifold.

The exchange of one massless scalar particle leads to an effective
 potential which is inversely proportional to  the separation
 distance.
The exchange of an even number of massless pseudoscalar particles
(for instance, by the arions) results in the effective potentials
inversely proportional to higher powers of the separation.
Similar potentials arise also due to the exchange of two
neutrinos,
two goldstinos, or other massless fermions \cite{29,30}.
The power-type effective potentials are also usually
parametrized as corrections to Newton's law of gravitation
\begin{equation}
V_n(|\mbox{\boldmath$r$}_1-\mbox{\boldmath$r$}_2|)=-
\frac{Gm_1m_2}{|\mbox{\boldmath$r$}_1-\mbox{\boldmath$r$}_2|}\,\left[1+
\Lambda_n
\left(\frac{r_0}{|\mbox{\boldmath$r$}_1-\mbox{\boldmath$r$}_2|}\right)^{n-1}
\right].
\label{eq5}
\end{equation}
\noindent
Here, $\Lambda_n$ is a dimensionless constant, $n$ is a positive
integer, and $r_0=10^{-15}\,$m is chosen to preserve the correct
dimension of energy at different $n$. Note that the exchange by
two axion-like particles in the limiting case $m_a\to 0$
in accordance to (\ref{eq3}) results in the potential \cite{31}
\begin{equation}
V_{kl}(|\mbox{\boldmath$r$}_1-\mbox{\boldmath$r$}_2|)=-
\frac{g_{ak}^2g_{al}^2}{64\pi^3m_km_l}\,
\frac{1}{|\mbox{\boldmath$r$}_1-\mbox{\boldmath$r$}_2|^3}.
\label{eq6}
\end{equation}
\noindent
This can be represented as a correction to Newton's law of
gravitation in (\ref{eq5}) with $n=3$ (the same power-type
interaction is obtained from the exchange of two arions).
The effective potential (\ref{eq5}) with $n=3$ is also obtained
from extra-dimensional models with noncompact (but warped)
extra dimensions \cite{32,33}.

\section{Constraints on an axion from measurements of the Casimir-Polder force}

The Casimir-Polder force acting between ${}^{87}$Rb atoms
belonging to a Bose-Einstein condensate cloud and a SiO${}_2$
plate was measured by means of the following dynamic
experiment \cite{34}. The condensate cloud was placed in a
magnetic trap with frequencies $\omega_{0z}=1438.85\,$rad/s
in the perpendicular direction to the plate and
$\omega_{0t}=40.21\,$rad/s in the lateral direction.
The Thomas-Fermi radii of the condensate cloud of ${}^{87}$Rb
atoms in the perpendicular and lateral directions were
$R_z=2.69\,\mu$m and $R_l=97.1\,\mu$m, respectively.
The dipole oscillations of the condensate in the $z$ direction
with a constant amplitude $A_z=2.5\,\mu$m were excited.
The separation distance $a$ between the center of mass of a
condensate and a plate was varied from 6.88 to $11\,\mu$m,
i.e., in the region where the thermal effects in the
Casimir-Polder force contribute essentially.
The temperature of the plate was equal to either $T=310\,$K
(as in an environment) or $T=479\,$K and $T=605\,$K (which
corresponds to out of equilibrium situations). However, for
constraining the parameters of an axion, the strongest result
follows from the measurements in thermal equilibrium.

Under the influence of the Casimir-Polder force between
${}^{87}$Rb atoms and a plate, the oscillation frequency
$\omega_{0z}$ slightly shifts to some other value $\omega_z$.
The relative frequency shift is given by
\begin{equation}
\gamma_z=\frac{|\omega_{0z}-\omega_z|}{\omega_{0z}}\approx
\frac{|\omega_{0z}^2-\omega_z^2|}{2\omega_{0z}^2}.
\label{eq7}
\end{equation}
\noindent
This frequency shift was measured \cite{34} as a function of
$a$ with some measurement errors determined at a 67\% confidence
level. For example, at the shortest separation $a_1=6.88\,\mu$m
this absolute error was $\Delta_1\gamma_z=3.06\times 10^{-5}$.
The quantity $\gamma_z$ was also calculated using the Lifshitz
theory of atom-wall interaction and subsequent averaging over the
condensate cloud. Under the assumption that SiO${}_2$ is an ideal
insulator, i.e., by disregarding the influence of its dc
conductivity,
it was found \cite{34} that the measurement results are in
agreement with theory in the limits of the experimental error
$\Delta\gamma_z$ (the importance of this assumption was
demonstrated later \cite{23,24,35}).

Due to the interaction potential (\ref{eq3}), there may be also
some additional force between a condensate cloud and a plate
caused by the two-axion exchange between protons and neutrons
belonging to them. The respective additional frequency shift can
be calculated by the additive summation of (\ref{eq3}) over all
nucleons of a ${}^{87}$Rb atom and a plate with subsequent
averaging over the condensate cloud (see \cite{12} for details).
Under an assumption that the plate has an infinitely large area
(it was shown \cite{12} that relative corrections to the result
due to a finite plate area are of order $10^{-6}$) the additional
frequency shift due to two-axion exchange is given by \cite{12}
\begin{equation}
\gamma_{z}^{\rm add}(a)=
\frac{15A(g_{ap},g_{an})}{2\pi A_z m_{\rm Rb}\omega_{0z}^2}
\Phi(a,m_a),
\label{eq8}
\end{equation}
\noindent
where $m_{\rm Rb}$ is the mass of ${}^{87}$Rb atom and the
function $\Phi(a,m_a)$ is defined as
\begin{eqnarray}
&&
\Phi(a,m_a)=
\int_{1}^{\infty}\!\!\!du\frac{\sqrt{u^2-1}}{u}
e^{-2m_aau}
\nonumber \\
&&~~~~~~\times
\left(1-e^{-2m_aDu}\right)\,
I_1(2m_aA_zu)\Theta(2m_aR_zu).
\label{eq9}
\end{eqnarray}
Here, $D=7\,$mm is the thickness of SiO${}_2$ plate and
\begin{equation}
\Theta(t)\equiv\frac{1}{t^3}(t^2\sinh t-3t\cosh t+3\sinh t).
\label{eq10}
\end{equation}
\noindent
The constant $A(g_{ap}g_{an})$ in (\ref{eq8}) depends on the
material properties as follows \cite{12}
\begin{eqnarray}
&&
A(g_{ap},g_{an})=
\frac{\rho_{{\rm SiO}_2}m_a}{16\pi^2m^2m_{\rm H}}
(37g_{ap}^2+50g_{an}^2)
\nonumber\\
&&~~~~~~~
\times\left(\frac{Z_{{\rm SiO}_2}}{\mu_{{\rm SiO}_2}}g_{ap}^2
+\frac{N_{{\rm SiO}_2}}{\mu_{{\rm SiO}_2}}g_{an}^2\right),
\label{eq11}
\end{eqnarray}
\noindent
where $\rho_{{\rm SiO}_2}$ is the plate density,
$m=(m_p+m_n)/2$ is the mean nucleon mass,
$Z_{{\rm SiO}_2}$ and $N_{{\rm SiO}_2}$ are the number of protons
and the mean number of neutrons in a SiO${}_2$ molecule,
respectively. The quantity
$\mu_{{\rm SiO}_2}=m_{{\rm SiO}_2}/m_{\rm H}$, where
$m_{{\rm SiO}_2}$ is the mean mass of a SiO${}_2$ molecule and
$m_{\rm H}$ is the mass of atomic hydrogen.

Taking into account that the observed frequency shift was in
agreement with that originating from the Casimir-Polder force,
the additional frequency shift (\ref{eq8}) due to two-axion
exchange should be constrained by the magnitude of the experimental
error
\begin{equation}
\gamma_{z}^{\rm add}(a_1)\leq\Delta_1\gamma_z.
\label{eq12}
\end{equation}
\noindent
{}From the numerical analysis of this equation, the constraints
on axion-nucleon coupling constants  were obtained \cite{12} under
different assumptions about a relationship between $g_{an}$ and
$g_{ap}$. For example, under a natural assumption that
$g_{an}=g_{ap}$ \cite{25}, the resulting constraints are shown
in Fig.~1, where the region of the plane above the line is
excluded and the region below the line is allowed.
These constraints cover the wide region of axion masses from
$m_a=10^{-4}$ to 0.3\,eV. As is seen in Fig.~1, the strength
of constraints decreases with increasing axion mass.
In Section 7 we compare the constraints of Fig.~1 with those
obtained from other measurements of the Casimir force and
different laboratory experiments.

\section{Constraints on an axion from measurements of the gradient of the Casimir force
by means of AFM}

In the sequence of three experiments, the gradient of the Casimir
force was measured
between the surfaces of a hollow sphere and a plate both coated
with Au films \cite{36,37},
with Au and Ni films, respectively \cite{38},
and with Ni films \cite{39,40}.
For technological purposes, there were also various material
layers
below Au and Ni coatings on both a hollow sphere made of fused
silica (SiO${}_2$) and a sapphire (Al${}_2$O${}_3$) plate.
The radii of spheres were of about $50\,\mu$m and the plates (disks)
were of approximately 5\,mm radius, i.e., by a factor of 100
larger than the spheres.
Measurements of the gradient of the Casimir force,
$\partial F_C(a)/\partial a$,
as a function of separation $a$ between the plate and the sphere,
were performed by means of dynamic AFM (see \cite{36,37} for
details). In all three experiments the measurement results were
found in agreement with theoretical predictions of the Lifshitz
theory in the limits of the experimental errors
$\Delta F_C^{\prime}(a)$. Calculations of the theoretical force
gradients were performed with omitted relaxation properties of
conduction electrons in metals (an account of the
relaxation properties of
conduction electrons in computations using the Lifshitz theory
leads to disagreement with the measurement data of many
experiments \cite{22,23,24,36,37,39,40}).

The two-axion exchange between nucleons belonging to a sphere and
a plate leads to some attraction in addition to the Casimir force.
 The gradient of this additional force acting between a spherical
envelope (layer) of thickness $\Delta_s$ and external radius $R$,
and a plate of thickness $D$ can be calculated by the additive
summation of the interaction potentials (\ref{eq3}) \cite{13}
\begin{eqnarray}
&&
\frac{\partial F_{\rm add}(a)}{\partial a}=
\frac{\pi}{m^2m_H^2}C_pC_s
\int_{1}^{\infty}\!du\frac{\sqrt{u^2-1}}{u^2}
\left(1-e^{-2m_auD}\right)
\nonumber \\
&&~~~~
\times
e^{-2m_aau}\,\left[\Phi(R,m_au)-
e^{-2m_au\Delta_s}\Phi(R-\Delta_s,m_au)\right],
\label{eq13}
\end{eqnarray}
\noindent
where the function $\Phi(r,z)$ is defined as
\begin{equation}
\Phi(r,z)=r-\frac{1}{2z}+e^{-2rz}\left(r+
\frac{1}{2z}\right),
\label{eq14}
\end{equation}
the coefficients $C_{p(s)}$ for a plate (spherical layer)
materials are given by
\begin{equation}
C_{p(s)}=\rho_{p(s)}\left(\frac{g_{ap}^2}{4\pi}\,
\frac{Z_{p(s)}}{\mu_{p(s)}}+\frac{g_{an}^2}{4\pi}\,
\frac{N_{p(s)}}{\mu_{p(s)}}\right),
\label{eq15}
\end{equation}
\noindent
$\rho_{p(s)}$ are the plate (spherical layer)
densities, and the quantities $Z_{p(s)}$, $N_{p(s)}$ and
$\mu_{p(s)}$ have the same meaning, as explained below
(\ref{eq11}), but in application to the molecules (atoms)
of a plate and a spherical layer, respectively.

Now we concentrate our attention on the experiment using
Au-coated surfaces of a spherical envelope of thickness
$\Delta_s^{\! g}=5\,\mu$m, of radius $R=41.3\,\mu$m
and a plate \cite{36,37}. The thicknesses of the Au coating
on the sphere and the plate were
$\Delta_s^{\!\rm Au}=\Delta_p^{\!\rm Au}=280\,$nm.
This allows to calculate the Casimir force (but not the
additive force due to two-axion exchange) as between entirely
Au bodies. In calculation of the additional force it should
be taken into account that in the experiment \cite{36,37}
the Au layers on both the spherical envelope and the plate
were deposited on the layers of Al of equal thicknesses
$\Delta_s^{\!\rm Al}=\Delta_p^{\!\rm Al}=20\,$nm.
Now the gradient of the additional force can be calculated
by applying (\ref{eq13}) to each pair of material layers
forming the spherical envelope and the plate taking into
account the separation distances between each pair of
material layers
\begin{eqnarray}
&&
\frac{\partial F_{\rm add}(a)}{\partial a}=
\frac{\pi}{m^2m_H^2}
\int_{1}^{\infty}\!du\frac{\sqrt{u^2-1}}{u^2}
e^{-2m_aau}
\nonumber \\
&&~~~~~~~~~~
\times
X_p(m_au)X_s(m_au),
\label{eq16}
\end{eqnarray}
\noindent
where
\begin{eqnarray}
&&
X_p(z)\equiv C_{\rm Au}\left(1-e^{-2z\Delta_p^{\!\rm Au}}\right)
\nonumber \\
&&~~~
+C_{\rm Al}e^{-2z\Delta_p^{\!\rm Au}}\left(1-e^{-2z\Delta_p^{\!\rm Al}}\right)
+C_{\rm sa}e^{-2z(\Delta_p^{\!\rm Au}+\Delta_p^{\!\rm Al})},
\nonumber \\[-2mm]
&&
\label{eq17} \\[-2mm]
&&
X_s(z)\equiv C_{\rm Au}\left[\Phi(R,z)-e^{-2z\Delta_s^{\!\rm Au}}
\Phi(R-\Delta_s^{\!\rm Au},z)\right]
\nonumber \\
&&~~~~~~
+C_{\rm Al}e^{-2z\Delta_s^{\!\rm Au}}
\left[\vphantom{e^{-2z\Delta_s^{\!\rm Al}}
\Phi(R-\Delta_s^{\!\rm Au})}
\Phi(R-\Delta_s^{\!\rm Au},z)\right.
\nonumber \\
&&~~~~~~~~~~\left.
-e^{-2z\Delta_s^{\!\rm Al}}
\Phi(R-\Delta_s^{\!\rm Au}-\Delta_s^{\!\rm Al},z)\right]
\nonumber \\
&&~~~~~~
+C_{g}e^{-2z(\Delta_s^{\!\rm Au}+\Delta_s^{\!\rm Al})}
\left[\vphantom{e^{-2z\Delta_s^{\!\rm Al}}
\Phi(R-\Delta_s^{\!\rm Au})}
\Phi(R-\Delta_s^{\!\rm Au}-\Delta_s^{\!\rm Al},z)\right.
\nonumber \\
&&~~~~~~~~~~\left.
-
e^{-2z\Delta_s^{\! g}}
\Phi(R-\Delta_s^{\!\rm Au}-\Delta_s^{\!\rm Al}
-\Delta_s^{\!g},z)\right].
\nonumber
\end{eqnarray}
\noindent
In these equations, the thickness of the sapphire plate
was put equal to
infinity, as it does not influence the result.
The coefficients $C_{\rm Au}$,  $C_{\rm Al}$,  $C_{g}$
and  $C_{sa}$ are defined in Eq.~(\ref{eq15}) which
should be applied to the
atoms Au and Al and to the molecules of glass and sapphire
[the densities of these materials entering (\ref{eq15}) are
$\rho_{\rm Au}$, $\rho_{\rm Al}$, $\rho_g$ and $\rho_{sa}$;
they can be found in the tables].

Taking into account that no additional force was observed in
the experiment \cite{36,37} within the measurement
error, one can write
\begin{equation}
\frac{\partial F_{\rm add}(a)}{\partial a}\leq
\Delta F_C^{\prime}(a).
\label{eq18}
\end{equation}
\noindent
Numerical analysis of this equation leads to new constraints
on the interaction constants $g_{ap}$ and $g_{an}$.
The strongest constraints are obtained at the shortest
experimental separation $a_1=235\,$nm. At this separation
distance the experimental error determined at a 67\% confidence
level is
$\Delta F_C^{\prime}(a_1)\equiv\Delta_1F_C^{\prime}=
0.5\,\mu$N/m \cite{36}.
In Fig.~2 we show these constraints by the solid line under
the assumption $g_{ap}=g_{an}$ (see \cite{13} for the alternative
assumptions). The region of the plane above the line is
excluded, and the region below the line is allowed.
The comparison of the solid line in Fig.~2 with the line in
Fig.~1 shows that the constraints following from measurements
of the gradient of the Casimir force are stronger than those
obtained from measurements of the Casimir-Polder force.
The largest strengthening by a factor of 170 is achieved for the
axion mass $m_a=0.3\,$eV.

Similar results can be obtained \cite{13} from the measurement
data of experiment with a Au-coated spherical envelope of
$R=64.1\,\mu$m radius and a Ni-coated plate \cite{38}.
The gradient of the additional force due to two-axion exchange
is again given by (\ref{eq16}), where $X_s(z)$ is presented in
(\ref{eq17}) and $X_p(z)$ takes a more simple form due to the
absence of an Al layer below a Ni coating
\begin{equation}
X_p(z)= C_{\rm Ni}\left(1-e^{-2z\Delta_p^{\!\rm Ni}}\right)
+C_{\rm Si}e^{-2z\Delta_p^{\!\rm Ni}}.
\label{eq19}
\end{equation}
Here, $\Delta_p^{\!\rm Ni}=154\,$nm and $C_{\rm Ni}$ can be
calculated using (\ref{eq15}).

The constraints on the coupling constants of axions to nucleons
can be again obtained from (\ref{eq18}).
The strongest constraints follow at the shortest separation
equal to $a_1=220\,$nm in this experiment. The respective total
experimental error determined at a 67\% confidence level is
$\Delta_1F_C^{\prime}=0.79\,\mu$N/m \cite{38}.
The constraints obtained under the condition $g_{ap}=g_{an}$
are shown by the long-dashed line in Fig.~2. As can be seen in
Fig.~2, the constraints following from the experiment with Au-Ni
test bodies are up to a factor 1.5 weaker
than those obtained
from the experiment with Au-Au test bodies. The main reason is
the smaller density of Ni, as compared with Au.

In the third experiment, a Ni-coated spherical envelope of
$R=61.71\,\mu$m radius and a Ni-coated plate were used \cite{39,40}.
The additional force can be again expressed by (\ref{eq16}).
In this case, however, the functions $X_p(z)$ and $X_s(z)$ are
more complicated than in the previously considered experiments
because for technological purposes there were two additional
layers (Al and Cr)
below the Ni coating on both a spherical envelope and on a plate
(see \cite{13} for explicit expressions).

The constraints on $g_{ap}=g_{an}$ were again obtained from (\ref{eq18}).
The strongest constraints follow at the shortest separation
distance ($a_1=223\,$nm in this case). The total
experimental error determined at a 67\% confidence level
at the shortest separation is
$\Delta_1F_C^{\prime}=1.2\,\mu$N/m \cite{38}.
The obtained constraints
are shown by the short-dashed line in Fig.~2. They are slightly
weaker than those following from the experiments with Au-Au
and Au-Ni test bodies. This is again explained by
the smaller density of Ni in comparison with that of Au (see
Section 7 for comparison with other laboratory constraints).

\section{Constraints on an axion from measurements of the  Casimir pressure
by means of micromachined oscillator}

The Casimir pressure $P_C(a)$ between two parallel Au-coated
plates
was determined from dynamic measurements performed in sphere-plate
 geometry using a  micromechanical torsional oscillator
\cite{41,42}.
A sapphire sphere and a Si plate of thickness $D=5\,\mu$m were
coated with the layers of Cr of equal thickness
$\Delta_s^{\!\rm Cr}=\Delta_p^{\!\rm Cr}=10\,$nm.
The outer layers of Au were of thicknesses
$\Delta_s^{\!\rm Au}=180\,$nm on the sphere and
$\Delta_p^{\!\rm Au}=210\,$nm on the plate.
The resulting radius of the sphere was measured to be
$R=151.3\,\mu$m. The experimental results for the Casimir pressure
 between two parallel plates spaced $a$ apart were found to be
in agreement with the predictions of the Lifshitz theory in the
limits of the total experimental error in the pressure
measurements $\Delta P_C(a)$ determined at a 95\% confidence
level. Here, we recalculate this error to a 67\% confidence
level in order to obtain constraints comparable with those
following from other experiments. The theoretical results were
obtained with omitted contribution of the relaxation properties
of free electrons (taking these properties into account leads
to theoretical predictions excluded by the measurement data
\cite{23,24,41,42}).

The additional effective pressure between two parallel plates
due to two-axion exchange between nucleons of a sphere and a
plate can be calculated by the additive summation using the
interaction potential (\ref{eq3}) (see \cite{14} for details).
The result is the following \cite{14}:
\begin{eqnarray}
&&
P_{\rm add}(a)=
-\frac{1}{2m^2m_{\rm H}^2R}\int_{1}^{\infty}\!\!\!du
\frac{\sqrt{u^2-1}}{u^2}
\nonumber \\
&&~~~~~~~~~
\times
e^{-2m_aau}\tilde{X}_p(m_au)\tilde{X}_s(m_au),
\label{eq20}
\end{eqnarray}
\noindent
where
\begin{eqnarray}
&&
\tilde{X}_p(z)\equiv C_{\rm Au}\left(1-e^{-2z\Delta_p^{\!\rm Au}}
\right)
\nonumber \\
&&~~~
+C_{\rm Cr}e^{-2z\Delta_p^{\!\rm Au}}
\left(1-e^{-2z\Delta_p^{\!\rm Cr}}
\right)
\nonumber \\
&&~~~
+C_{\rm Si}e^{-2z(\Delta_p^{\!\rm Au}+\Delta_p^{\!\rm Cr})}
\left(1-e^{-2zD}
\right),
\label{eq21} \\[1mm]
&&
\tilde{X}_s(z)\equiv C_{\rm Au}\left[
\vphantom{e^{-2z\Delta_s^{\!\rm Au}}}
\Phi(R,z)
-e^{-2z\Delta_s^{\!\rm Au}}
\Phi(R-\Delta_s^{\!\rm Au},z)\right]
\nonumber \\
&&~~~
+C_{\rm Cr}e^{-2z\Delta_s^{\!\rm Au}}
\left[
\vphantom{e^{-2m_au\Delta_s^{\!\rm Au}}}
\Phi(R-\Delta_s^{\!\rm Au},z)
\right.
\nonumber \\
&&~~~~~~~~~~~~
\left.
-e^{-2z\Delta_s^{\!\rm Cr}}
\Phi(R-\Delta_s^{\!\rm Au}-\Delta_s^{\!\rm Cr},z)
\right]
\nonumber \\
&&~~~
+C_{sa}
e^{-2z(\Delta_s^{\!\rm Au}+\Delta_s^{\!\rm Cr})}
\Phi(R-\Delta_s^{\!\rm Au}-\Delta_s^{\!\rm Cr},z).
\nonumber
\end{eqnarray}
\noindent
The function $\Phi(r,z)$ used here is given in (\ref{eq14}).
The coefficients $C_{\rm Au}$, $C_{\rm Cr}$, $C_{\rm Si}$, and
$C_{sa}$ are the same as used above. All of them are expressed
by (\ref{eq15}), as applied to respective materials.

The constraints on the axion-nucleon interaction constants were
found from the inequality
\begin{equation}
|P_{\rm add}(a)|\leq\Delta P_C(a).
\label{eq22}
\end{equation}
\noindent
For different regions of axion masses the strongest constraints
follow from (\ref{eq22}) at different separation distances.
Thus, within the regions
$m_a<0.1\,$eV, $0.1\,\mbox{eV}\leq m_a<0.5\,$eV and
$0.5\,\mbox{eV}\leq m_a<15\,$eV the strongest constraints were
obtained at $a=300$, 200 and 162\,nm, respectively.
At these separations the total experimental errors in
measurements of the Casimir pressure recalculated to a 67\%
confidence level were equal to 0.22, 0.38, and 0.55\,mPa,
respectively. In Fig.~3 the obtained constraints are shown
by the solid line under the condition $g_{ap}=g_{an}$.
They are stronger than the constraints following from
measurements of the Casimir-Polder force (see Fig.~1) and from
measurements of the gradient of the Casimir force between
Au-Au surfaces (see the solid line in Fig.~2). Thus, at
$m_a=1\,$eV
the constraints of Fig.~3 are stronger by a factor of 3.2 than
the strongest constraints of Fig.~2 shown by the solid line
(a more detailed comparison is contained in Section 7).

\section{Constraints on an axion from measurements of the  Casimir force
between corrugated surfaces}

Several measurements of the Casimir interaction between a sphere
and a plate were performed in the case when the surface of at
 least one test body is not smooth, but covered with the
 longitudinal corrugations \cite{43,44,45,46,47,48,49,50}.
 The shape of the corrugations was either sinusoidal
 \cite{43,44,47,48,49,50} or rectangular \cite{45,46}
(in the latter case the sphere was smooth, and only the plate was
corrugated). If both the test bodies are  corrugated and some
nonzero phase shift between corrugations is present, there is
not only the normal Casimir force acting perpendicular to the
surfaces, but the lateral Casimir force as well
\cite{43,44,47,48}.
Here we consider the constraints on axion-nucleon coupling
constants obtained \cite{15} from measurements of the normal
\cite{49,50} and lateral \cite{47,48} Casimir force between
sinusoidally corrugated Au-coated surfaces (experiments
\cite{43,44} are less precise, and experiments \cite{45,46}
use the rectangular corrugated Si plates and lead to weaker
constraints due to a smaller density of Si).

We begin with an experiment on measuring the lateral Casimir
force  between
sinusoidally corrugated surfaces of a sphere and a plate
\cite{47,48}. The corrugation axes of the longitudinal
corrugations on both bodies were kept parallel, and there was
some phase shift $\varphi_0$ between corrugations.
The period of corrugations was $\Lambda=574.4\,$nm.
Measurements of the lateral Casimir force as a function of
the phase shift were performed over the region of separations
between the mean levels of corrugations from 120 to 190\,nm.
The corrugation amplitudes were
$A_1=85.4\,$nm and $A_2=13.7\,$nm on the plate and on the
sphere, respectively. The plate was made of a hard epoxy and
coated with a layer of Au of thickness
$\Delta_p^{\!\rm Au}=300\,$nm.
The sphere was made of polystyrene and coated with a layer of
Cr of $\Delta_s^{\!\rm Cr}=10\,$nm thickness and then with a layer of
Au of $\Delta_s^{\!\rm Au}=50\,$nm thickness.
The outer radius of the sphere was measured to be $R=97.0\,\mu$m.
The measurement results were compared with theoretical predictions
 of the scattering theory (which generalizes the Lifshitz theory
 for the case of arbitrary shaped bodies) and demonstrated good
 agreement in the limits of the experimental error
 $\Delta F_C^{\rm lat}(a)$ \cite{47,48}.

 The additional lateral force due to two-axion exchange between
 sinusoidally corrugated surfaces of a sphere and a plate can be
 calculated using (\ref{eq3}). The maximum amplitude of this
 force, which is obtained at the phase shift $\varphi_0=\pi/2$,
 takes the form \cite{15}
\begin{eqnarray}
&&
\max|F_{\rm add}^{\rm lat}(a)|=
\frac{\pi^2 RC_{\rm Au}}{m_am^2m_{\rm H}^2}\,
\frac{A_1A_2}{\Lambda\sqrt{A_1^2+A_2^2}}
\nonumber \\[1mm]
&&~~
\times
\int_{1}^{\infty}\!\!\!du\frac{\sqrt{u^2-1}}{u^3}
e^{-2m_aua} I_1\left(2m_au\sqrt{A_1^2+A_2^2}\right)
\nonumber \\[1mm]
&&~~~~~
\times
(1-e^{-2m_au\Delta_p^{\!\rm Au}})
\left[
\vphantom{e^{-2m_au\Delta_{\rm Au}^{\!(1)}}}
C_{\rm Au}+(C_{\rm Cr}-C_{\rm Au})
\right.
\nonumber \\[1mm]
&&~~~~\left.
\times e^{-2m_au\Delta_s^{\!\rm Au}}
-C_{\rm Cr}
e^{-2m_au(\Delta_s^{\!\rm Au}+\Delta_s^{\!\rm Cr})}
\right].
\label{eq23}
\end{eqnarray}
\noindent
Here, the hard epoxy and polystyrene would lead to negligibly
small contributions to the force due to two-axion exchange.
Because of this, only metallic coatings were taken into account
in (\ref{eq23}).

The constraints on an axion can be obtained from the inequality
\begin{equation}
\max|F_{\rm add}^{\rm lat}(a)|\leq
\Delta F_{C}^{\rm lat}(a),
\label{eq24}
\end{equation}
\noindent
where the left-hand side is given by (\ref{eq23}).
For axion-like particles with masses $m_a<20\,$eV, the strongest
constraints are obtained from the measure of agreement between
experiment and theory at $a=124.7\,$nm. At this separation the
total experimental error recalculated to a 67\% confidence level
for convenience in comparison with other experiments is
$\Delta F_C^{\rm lat}=2.4\,$pN (note that according to a
conservative estimation, the total experimental error
calculated in \cite{47,48} at a 95\% confidence level is by a
factor of 2 larger than the same error found at a 67\% confidence
level). The constraints on $g_{ap}=g_{an}$ obtained from
(\ref{eq24}) at $a=124.7\,$nm
are shown by the solid line in Fig.~4, where the region of the
plane above the line is excluded and the region below the line is
allowed. Note that this line is slightly different from the
respective lines in Fig.~2(a,b) in \cite{15} because it was
plotted there at the 95\% confidence level.

We now turn our attention to the experiment on measuring the
normal Casimir force between a sinusoidally corrugated Au-coated
polystyrene sphere of $R=99.6\,\mu$m radius and a sinusoidally
corrugated Au-coated plate made of hard epoxy \cite{49,50}.
This experiment was performed at different angles between the
longitudinal corrugations on the sphere and on the plate varying
from 0 to 2.4${}^{\circ}$. There was no phase shift between
corrugations on both bodies. Below we obtain constraints on the
axion-nucleon coupling constants from the measurement data for
the case of parallel corrugation axes on the sphere and the
plate.  The thicknesses of Au coatings on the sphere and on the
plate were $\Delta_s^{\!\rm Au}=110\,$nm and
$\Delta_p^{\!\rm Au}=300\,$nm, respectively.
For technological purposes, before depositing the Au coatings,
the sphere was first coated with a layer of Cr of thickness
$\Delta_s^{\!\rm Cr}=10\,$nm and then
with a layer of Al of thickness
$\Delta_s^{\!\rm Al}=20\,$nm.
The period of uniaxial sinusoidal corrugations on both bodies
was $\Lambda=570.5\,$nm, and the corrugations amplitudes were
$A_1=40.2\,$nm and $A_2=14.6\,$nm on the plate and on the sphere,
respectively. The measurement results were compared with
theoretical predictions of the scattering theory and found in
good agreement within the limits of the total experimental error.

The additional normal force acting between a sphere and a plate
due to two-axion exchange was again calculated \cite{15} using
(\ref{eq3})
\begin{eqnarray}
&&
F_{\rm add}^{\rm nor}(a)=-\frac{\pi RC_{\rm Au}}{2m_am^2m_{\rm H}^2}
\int_{1}^{\infty}\!\!\!du\frac{\sqrt{u^2-1}}{u^3}e^{-2m_aua}
\nonumber \\[1mm]
&&~~~~~
\times  I_0\left(2m_au(A_1-A_2)\right)(1-e^{-2m_au\Delta_p^{\,\rm Au}})
\nonumber \\[1mm]
&&~~~~~
\times\left[C_{\rm Au}+(C_{\rm Al}-C_{\rm Au})
e^{-2m_au\Delta_s^{\!\rm Au}}\right.
\nonumber \\[1mm]
&&~~~~~~~
+(C_{\rm Cr}-C_{\rm Al})
e^{-2m_au(\Delta_s^{\!\rm Au}+\Delta_s^{\!\rm Al})}
\nonumber \\[1mm]
&&~~~~~~~
\left.
-C_{\rm Cr}
e^{-2m_au(\Delta_s^{\!\rm Au}+\Delta_s^{\!\rm Al}
+\Delta_s^{\!\rm Cr})}\right].
\label{eq25}
\end{eqnarray}
\noindent
The constraints on the axion-nucleon coupling constants
$g_{an}=g_{ap}$ were found from the inequality
\begin{equation}
|F_{\rm add}^{\rm nor}(a)|\leq\Delta F_C^{\rm nor}(a).
\label{eq26}
\end{equation}
\noindent
The strongest constraints follow from (\ref{eq26}) at the
shortest separation distance $a_1=127\,$nm where the total
experimental error determined at a 67\% confidence level is
equal to $\Delta F_C^{\rm nor}(a_1)=0.94\,$pN \cite{49,50}.

In Fig.~4 the obtained constraints under a condition
$g_{an}=g_{ap}$ are shown by the dashed line. It can be seen
that for $m_a<5.3\,$eV they are stronger than those following
from measurements of the lateral Casimir force (the solid line),
but become weaker than the latter for larger axion masses.

\section{Comparison between different laboratory constraints}

It is interesting to compare all discussed above constraints,
obtained from measurements of the Casimir interaction,
between themselves and with other laboratory constraints on
axion to nucleon coupling constants. Such a comparison is
performed in Fig.~5 over the wide range of axion masses
from $10^{-10}$ to 20\,eV. The constraints on $g_{an}$
obtained \cite{51} by means of a magnetometer using
spin-polarized K and ${}^3$He atoms are shown by the solid
line 1. These constraints are applicable in the region of
$m_a$ from $10^{-10}$ to $6\times 10^{-6}\,$eV.
The solid line 2 indicates the constraints obtained \cite{52}
from the recent Cavendish-type experiment \cite{53} in the
region from $m_a=10^{-6}$ to $6\times 10^{-2}\,$eV.
The weaker constraints found \cite{25} from the older
Cavendish-type experiments \cite{54,55} and from the
E\"{o}tvos-type experiment \cite{56}, respectively, are
shown by the dashed lines 3 and 4 (these and the following
constraints are obtained under a condition $g_{an}=g_{ap}$).
These constraints cover the region of $m_a$ from $10^{-8}\,$eV
to $4\times 10^{-5}\,$eV (line 3) and to $10^{-5}\,$eV (line 4).
The lines 5--8 are obtained \cite{12,13,14,15} from measurements
of the Casimir interaction. They are discussed in this paper.
The line 5 reproduces the line in Fig.~3 obtained for $m_a$
from $10^{-3}$ to 15\,eV from measurements of the Casimir
pressure (see Section 5). The dashed lines 6 and 7 reproduce
the solid line in Fig.~2 and the line in Fig.~1 found in the
region from $3\times 10^{-5}$ to 1\,eV from measurements of
the gradient of the Casimir force between Au-Au surfaces and
in the region from $10^{-4}$ to 0.3\,eV from measurements of
the Casimir-Polder force, respectively (see Sections 4 and 3).
Finally, the line 8 reproduces the solid line in Fig.~4
found in the region of $m_a$ from 1 to 20\,eV.
It follows from measurements of the lateral Casimir force
between corrugated surfaces discussed in Section 6
(measurements of the normal Casimir force between sinusoidally
corrugated surfaces lead to weaker constraints than those
shown in Fig.~5).

The strength of almost all laboratory constraints shown in
Fig~5 (with exception of that shown by line 1) monotonically
decreases with increase of the axion mass $m_a$.
If one introduces the Compton wavelength of an axion
$\lambda_a=1/m_a$, it is correct to say that the strength
of almost all constraints (and all of those
following from measurements
of the gravitational and Casimir interactions) decreases with
decreasing $\lambda_a$. The same is true for the
Yukawa-type corrections to Newton's law of gravitation
(\ref{eq4}) whose strength decreases with decreasing
interaction range $\lambda$ (see the next section).
This property likens the interaction potentials (\ref{eq3})
and (\ref{eq4}) and specifies the interaction range where
the most strong constraints on respective hypothetical
forces can be obtained from experiments on measuring the
Casimir interaction.

The vertical lines in Fig.~5 indicate the region from
$m_a=10^{-5}$ to $10^{-2}\,$eV, which is often called an
axion window \cite{57}. As can be seen in Fig.~5,
experiments measuring the Casimir interaction lead to
strengthening of the laboratory constraints on axion to
nucleon coupling constants near the upper border of the
axion window and also for larger axion masses.

\section{Constraints on corrections to Newton's law of gravitation}

The constraints on corrections to the Newton law of gravitation
described by the potentials (\ref{eq4}) and (\ref{eq5}) can be
obtained from the gravitational experiments of E\"{o}tvos- and
Cavendish-type and from measurements of the Casimir interaction.
As explained in Section 1, measurements of the Casimir force
have long been used for constraining hypothetical interactions
of both Yukawa and power type.
Because of this, here we only briefly present the obtained
results and indicate regions where measurements of the Casimir
force lead to the most strong constraints, as compared to
gravitational experiments.

The Yukawa-type interaction potential between the test bodies
used in experiments on measuring the Casimir force is obtained
by the integration of (\ref{eq4}) over the volumes of bodies.
In so doing, at submicrometer separations the Newton
gravitational force turns out to be negligibly small, as
compared to the error of force measurements. Similar to the
case of axion considered above, the constraints on the constants
of Yukawa-type interaction $\alpha$ and $\lambda$ are obtained
from a condition that this interaction was not experimentally
observed in the limits of the experimental error in measurements
of the Casimir interaction.

In Fig.~6 we present the strongest constraints on the
Yukawa interaction constant $\alpha$ in the micrometer and
submicrometer interaction range $\lambda$ obtained
from measurements of the Casimir interaction.
The line 1 in Fig.~6 was obtained \cite{18} from measurements
of the lateral Casimir force between sinusoidally corrugated
surfaces of a sphere and a plate \cite{47,48} (see Section 6).
It presents the strongest constraints on the Yukawa-type
corrections to Newton's law of gravitation within the
interaction range from $\lambda=1.6$ to 11.6\,nm.
The line 2 shows constraints found \cite{21} from measuring
the normal Casimir force between sinusoidally corrugated
surfaces at the angle between corrugations equal to
2.4${}^{\circ}$ \cite{49,50} (see Section 6). These constraints
are the strongest ones in the interaction range from 11.6 to
17.2\,nm.
The constraints obtained from measurements of the Casimir
pressure by means of a micromachined torsional oscillator
(see Section 5) are
indicated by the line 3. They are the strongest ones for
$17.2\,\mbox{nm}<\lambda<89\,$nm.
At larger $\lambda$ the most strong constraints shown by the
line 4 follow from the so-called Casimir-less experiment
\cite{58}, where the Casimir force was nullified by using the
difference force measurement scheme.  These constraints are
the strongest ones up to $\lambda=891\,$nm.
The constraints of the line 5 are found \cite{59} from
measurements of the Casimir force between Au-coated surfaces
of a plate and a spherical lens of large radius. They are
the strongest ones up to $\lambda=3.16\,\mu$m.
For larger $\lambda$ the strongest constraints on the
Yukawa-type corrections to Newton's gravitational law
follow from the Cavendish-type experiments. The first
constraints of such kind are indicated by the line 6
\cite{60,61}. Thus, measurements of the Casimir interaction
lead to the most strong constraints on non-Newtonian
gravity over a wide interaction range from 1.6\,nm to a
few micrometers. As can be seen in Fig.~6, the strength of
all constraints decreases with decreasing $\lambda$, i.e.,
with increasing mass of a hypothetical particle which
initiates the additional interaction of Yukawa-type.
This is similar to the case of an axion considered in
Sections 3--6.

Constraints on the power-type corrections to Newton's law
of gravitation (\ref{eq5}) follow from the gravitational
experiments of E\"{o}tvos and Cavendish type \cite{8} and
from measurements of the Casimir force \cite{17,30}.
At the present time the most strong constraints follow from
the E\"{o}tvos-type experiments
($|\Lambda_1|\leq 1\times 10^{-9}$ \cite{62} and
$|\Lambda_2|\leq 4\times 10^{8}$ \cite{56}) and from
the Cavendish-type experiments
($|\Lambda_3|\leq 1.3\times 10^{20}$ \cite{52},
$|\Lambda_4|\leq 4.9\times 10^{31}$ \cite{52}, and
$|\Lambda_5|\leq 1.5\times 10^{43}$ \cite{52}).
Note that \cite{52} uses another parametrization for the
power-type corrections to Newtonian gravitation.

\section{Conclusions and discussion}
In the foregoing, we have considered the constraints on axion
to nucleon couplings following from laboratory experiments on
measuring the Casimir interaction. The obtained constraints
are quite competitive in the region of axion masses from
$10^{-3}$ to 20\,eV. The most strong of them follow from a
dynamic determination of the Casimir pressure between two
parallel plates and from measurement of the lateral Casimir
force between sinusoidally corrugated surfaces.
All these constraints were derived by considering the process
of two-axion exchange between two nucleons. This process is of
the lowest order contributing to the force acting between
unpolarized test bodies. The obtained constraints were
compared with those following from other laboratory experiments.

We have also compared the constraints on an axion with previously
obtained constraints on corrections to the Newton law of
gravitation of Yukawa and power type. The most strong constraints
of this kind following from measurements of the Casimir
interaction are collected.
In the interaction range below a few micrometers they are
stronger than the constraints on Yukawa-type corrections to
Newton's law following from the gravitational experiments of
E\"{o}tvos and Cavendish type.

In future it would be interesting to perform measurements of
the Casimir interaction between two polarized test bodies.
This would lead to an additional force due to exchange of one
axion between protons and neutrons and, as a consequence, to
much stronger constraints on the axion to nucleon coupling
constants.


\begin{figure}[t]
\vspace*{-15cm}
\centerline{\hspace*{2.5cm}
\includegraphics{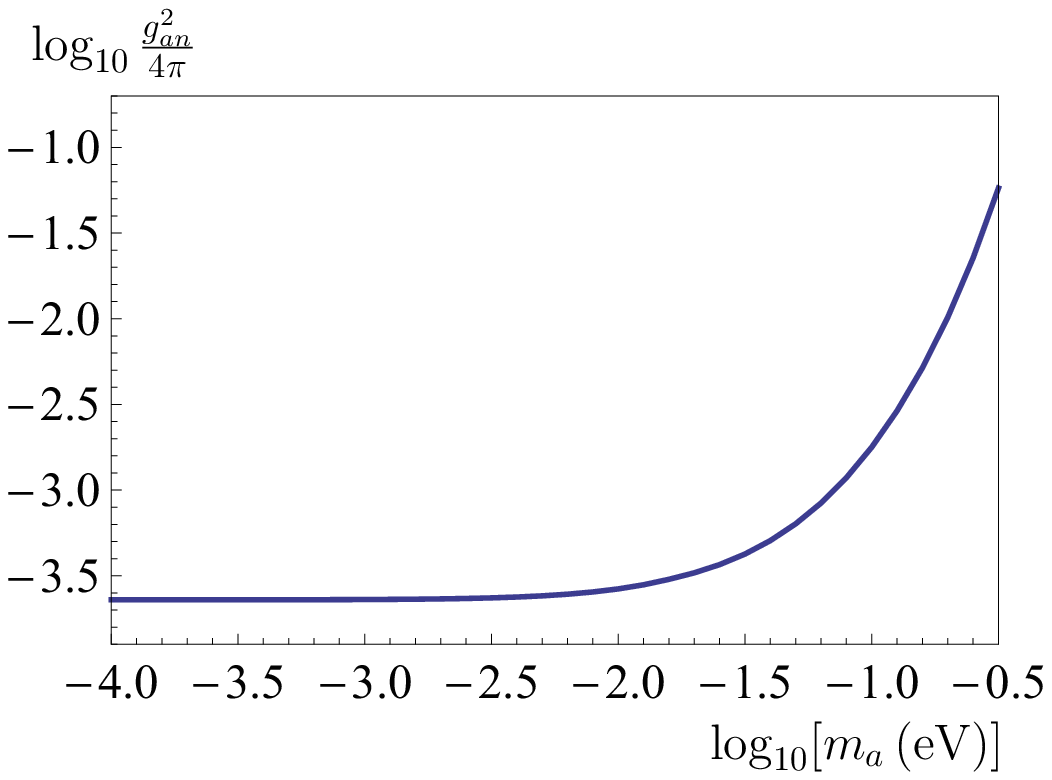}
}
\vspace*{-5cm}
\caption{Constraints on the coupling constants
of an axion with a proton and a neutron following from
measurements of the thermal Casimir-Polder force are shown
as a function of the axion mass.
 The region of the plane above the line
is prohibited and below the line is allowed.
}
\end{figure}
\begin{figure}[b]
\vspace*{-15cm}
\centerline{\hspace*{2.5cm}
\includegraphics{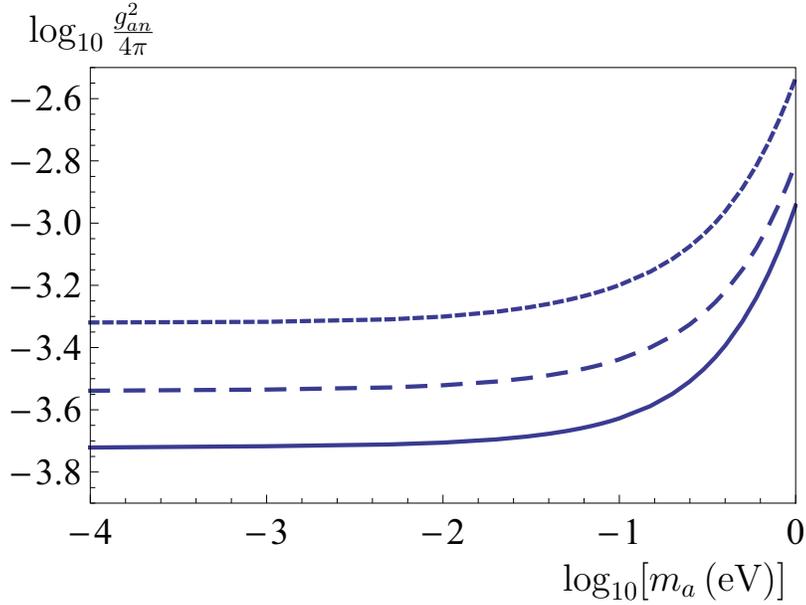}
}
\vspace*{-5cm}
\caption{Constraints on the coupling constants
of an axion with a proton and a neutron following from
measurements of the gradient of the Casimir force
between Au-Au, Au-Ni and Ni-Ni surfaces are shown
as  functions of the axion mass by the solid, long-dashed
and short-dashed lines, respectively.
 The regions of the plane above the lines
are prohibited and below the lines are allowed.
}
\end{figure}
\begin{figure}[b]
\vspace*{-15cm}
\centerline{\hspace*{2.5cm}
\includegraphics{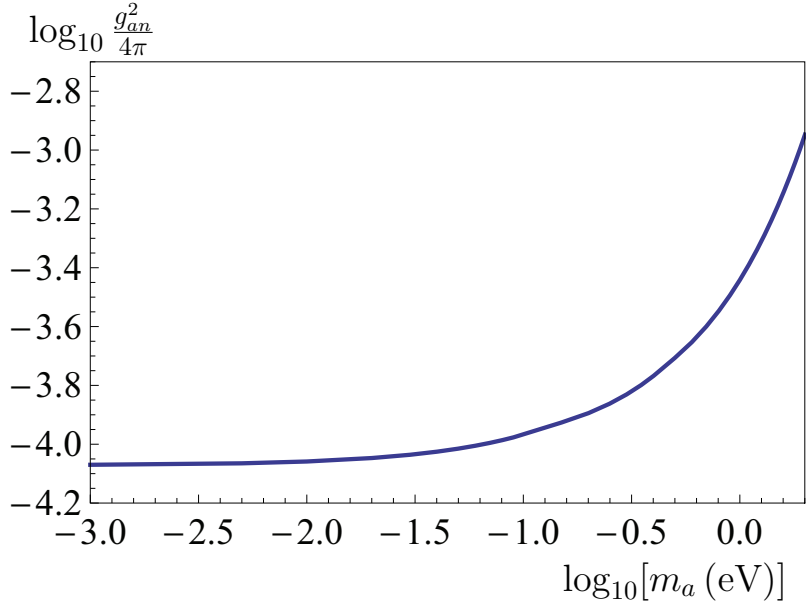}
}
\vspace*{-5cm}
\caption{Constraints on the coupling constants
of an axion with a proton and a neutron following from
dynamic determination of the Casimir pressure
between two parallel Au plates are shown
as a function of the axion mass.
 The region of the plane above the line
is prohibited and below the line is allowed.
}
\end{figure}
\begin{figure}[b]
\vspace*{-15cm}
\centerline{\hspace*{2.5cm}
\includegraphics{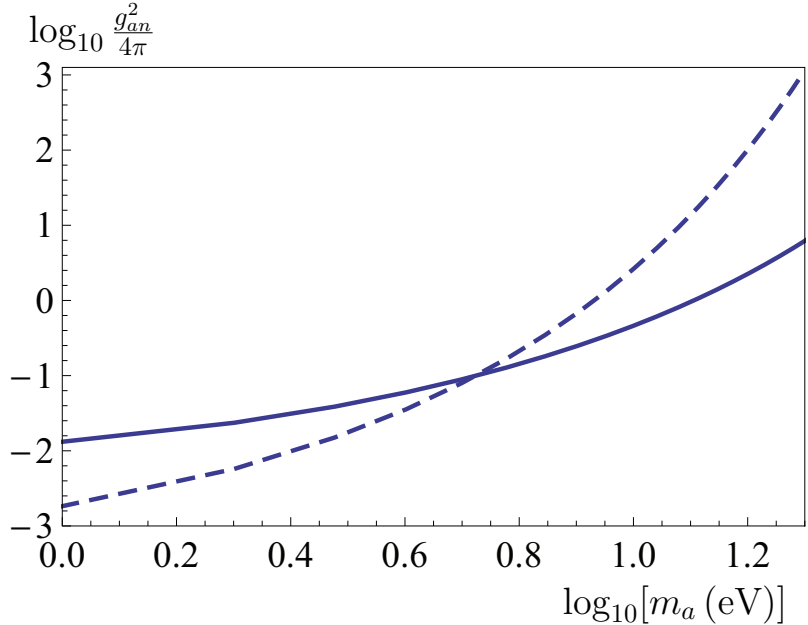}
}
\vspace*{-5cm}
\caption{Constraints on the coupling constants
of an axion with a proton and a neutron following from
measurements of the lateral (the solid line) and
normal (the dashed line) Casimir forces between
sinusoidally corrugated surfaces are shown
as functions of the axion mass.
 The regions of the plane above the lines
are prohibited and below the lines are allowed.
}
\end{figure}
\begin{figure}[b]
\vspace*{-8cm}
\centerline{\hspace*{2.5cm}
\includegraphics{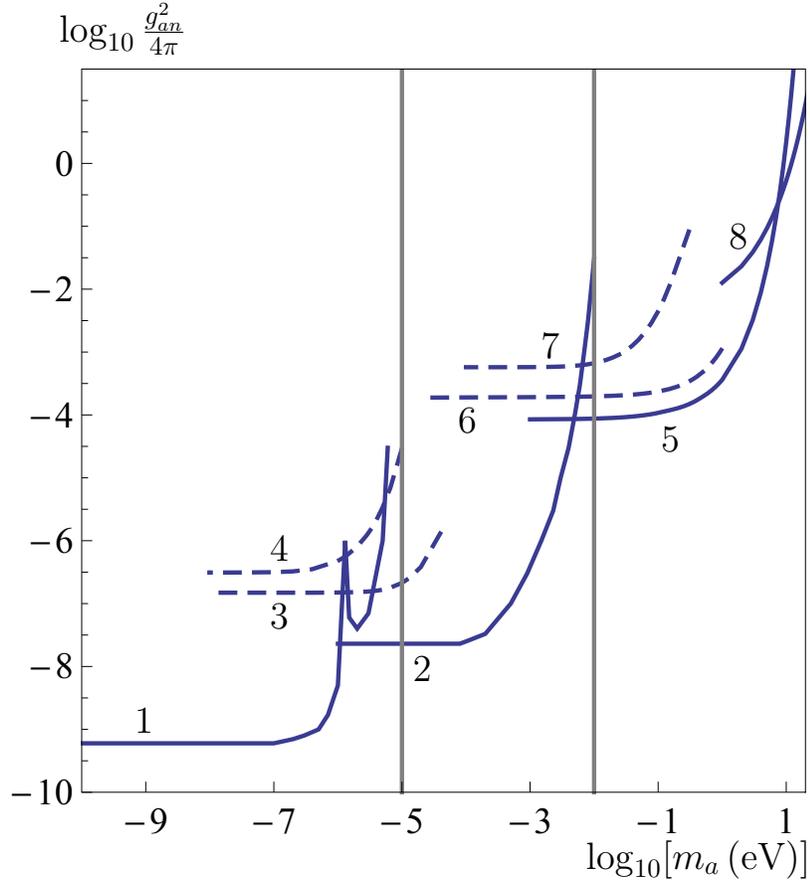}
}
\vspace*{-7cm}
\caption{Different laboratory constraints on the coupling constant
of an axion with a neutron following from
the magnetometer measurements (the line 1), from the
Cavendish- and E\"{o}tvos-type experiments (the lines 2--4),
from measurements of the Casimir pressure (the line 5),
of the gradient of the Casimir force (the line 6),
of the Casimir-Polder force (the line 7), and of the
lateral Casimir force (the line 8) are shown
as functions of the axion mass.
The two vertical lines indicate the borders of the axion
window.
 The regions above each line
are prohibited and below each line are allowed.
}
\end{figure}
\begin{figure}[b]
\vspace*{-3cm}
\centerline{\hspace*{2.5cm}
\includegraphics{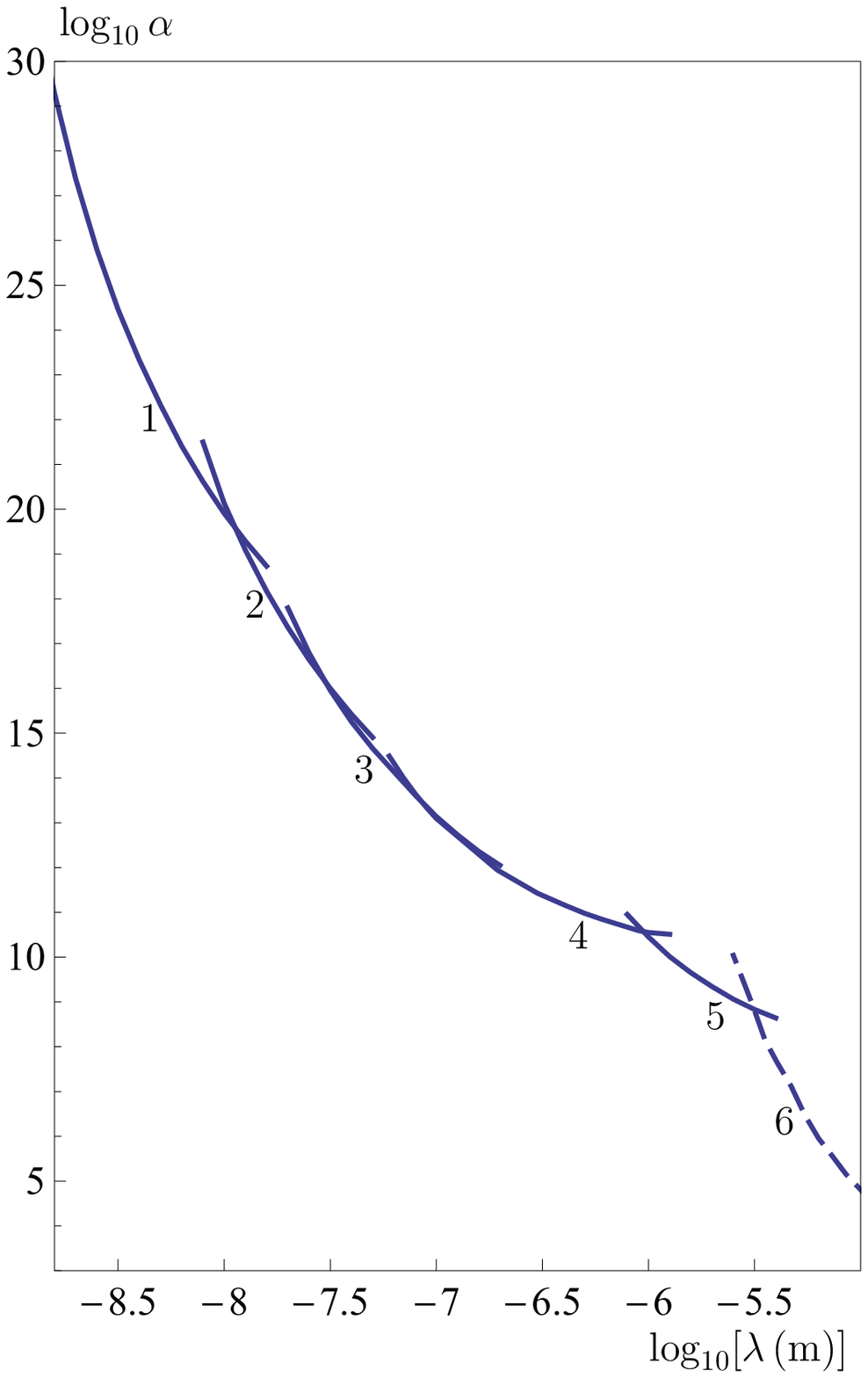}
}
\vspace*{-8cm}
\caption{Constraints on the Yukawa-type corrections to
the Newton law of gravitation obtained
from measurements of the lateral Casimir force between
corrugated surfaces (the line 1),
of the normal Casimir force between
corrugated surfaces (the line 2),
of the Casimir pressure (the line 3),
from the Casimir-less experiment (the line 4),
from measurements  of the Casimir force
between a plate and a spherical lens (the line 5),
and from the Cavendish-type experiment (the line 6)
are shown
as functions of the interaction range.
 The regions of the plane above the lines
are prohibited and below the lines are allowed.
}
\end{figure}

\begin{thebibliography}{99}
\bibitem{1}
S.~Weinberg,
Phys. Rev. Lett. {\bf 40}, 223 (1978).
\bibitem{2}
F.~Wilczek,
Phys. Rev. Lett. {\bf 40}, 279 (1978).
\bibitem{3}
R.~D.~Peccei and H.~R.~Quinn,
Phys. Rev. Lett. {\bf 38}, 1440 (1977).
\bibitem{4}
J.~E.~Kim,
Phys. Rep. {\bf 150}, 1 (1987).
\bibitem{5}
Yu.~N.~Gnedin,
Int. J. Mod. Phys. A {\bf 17}, 4251 (2002).
\bibitem{6}
K.~Baker {\it et al.}
Ann. Phys. (Berlin) {\bf 525}, A93 (2013).
\bibitem{7}
S.~Dimopoulos and G.~F.~Giudice,
 Phys. Lett. B {\bf 379}, 105 (1996).
\bibitem{8}
E.~Fischbach and C.~L.~Talmadge, {\it The Search for Non-Newtonian
Gravity} (Springer, New York, 1999).
\bibitem{9}
I.~Antoniadis,
N.~Arkani-Hamed, S.~Dimopoulos, and G.~Dvali,
Phys. Lett. B {\bf 436}, 257 (1998).
\bibitem{10}
N.~Arkani-Hamed, S.~Dimopoulos, and G.~Dvali,
Phys. Rev. D {\bf 59}, 086004 (1999).
\bibitem{11}
E.~G.~Adelberger, J.~H.~Gundlach, B.~R.~Heckel, S.~Hoedl, and
S.\ Schlamminger,
{Part. Nucl. Phys. } {\bf 62}, 102 (2009).
\bibitem{12}
V.~B.~Bezerra, G.~L.~Klimchitskaya,
 V.~M.~Mostepanenko, and C.~Romero,
Phys. Rev. D {\bf 89}, 035010 (2014).
\bibitem{13}
V.~B.~Bezerra, G.~L.~Klimchitskaya,
 V.~M.~Mostepanenko, and C.~Romero,
Phys. Rev. D {\bf 89}, 075002 (2014).
\bibitem{14}
V.~B.~Bezerra, G.~L.~Klimchitskaya,
 V.~M.~Mostepanenko, and C.~Romero,
Eur. Phys. J. C {\bf 74}, 2859 (2014).
\bibitem{15}
V.~B.~Bezerra, G.~L.~Klimchitskaya,
 V.~M.~Mostepanenko, and C.~Romero,
Phys. Rev. D {\bf 90}, 055013 (2014).
\bibitem{16}
V.~A.~Kuzmin, I.~I.~Tkachev, and M.~E.~Shaposhnikov,
Pis'ma v ZhETF {\bf 36}, 49 (1982)
[JETP Lett. {\bf 36}, 59 (1982)].
\bibitem{17}
V.\ M.\ Mostepanenko and I.~Yu.~Sokolov,
Phys. Lett. A {\bf 125}, 405 (1987).
\bibitem{18}
V.~B.~Bezerra, G.~L.~Klimchitskaya,
 V.~M.~Mostepanenko, and C.~Romero,
Phys. Rev. D {\bf 81}, 055003 (2010).
\bibitem{19}
V.~B.~Bezerra, G.~L.~Klimchitskaya,
 V.~M.~Mostepanenko, and C.~Romero,
Phys. Rev. D {\bf 83}, 075004 (2011).
\bibitem{20}
G.~L.~Klimchitskaya, U.~Mohideen, and
V.\ M.\ Mos\-te\-pa\-nen\-ko,
Phys. Rev. D {\bf 86}, 065025 (2012).
\bibitem{21}
G.~L.~Klimchitskaya, U.~Mohideen, and
V.\ M.\ Mos\-te\-pa\-nen\-ko,
Phys. Rev. D {\bf 87}, 125031 (2013).
\bibitem{22}
G.~L.~Klimchitskaya and
V.\ M.\ Mos\-te\-pa\-nen\-ko,
Grav. Cosmol. {\bf 20}, 3 (2014).
\bibitem{23}
M.~Bordag, G.~L.~Klimchitskaya, U.\ Mohideen, and
V.\ M.\ Mostepanenko, {\it Advances in the Casimir Effect}
(Oxford University Press, Oxford, 2009).
\bibitem {24}
G.~L.~Klimchitskaya, U. Mohideen, and V.\ M.\ Mostepanenko,
Rev. Mod. Phys. {\bf 81}, 1827 (2009).
\bibitem{25}
E.~G.~Adelberger, E.~Fischbach, D.~E.~Krause, and R.\ D.\ Newman,
{Phys. Rev. D} {\bf 68}, 062002 (2003).
\bibitem{26}
A.~Bohr and B.~R.~Mottelson,
{\it Nuclear Structure}
(Benjamin, New York, 1969), Vol.~1.
\bibitem{27}
S.~D.~Drell and K.~Huang,
Phys. Rev. {\bf 91}, 1527 (1953).
\bibitem{28}
F.~Ferrer and M.~Nowakowski,
Phys. Rev. D {\bf 59}, 075009 (1999).
\bibitem{29}
E.~Fischbach,
Ann. Phys. (N.Y.)  {\bf 247}, 213 (1996).
\bibitem{30}
V.\ M.\ Mostepanenko and I.~Yu.~Sokolov,
Phys. Rev. D  {\bf 47}, 2882 (1993).
\bibitem{31}
V.\ M.\ Mostepanenko and I.~Yu.~Sokolov,
Sov. J. Nucl. Phys.  {\bf 46}, 685 (1987).
\bibitem{32}
L.~Randall and R.~Sundrum,
Phys. Rev. Lett. {\bf 83}, 3370 (1999).
\bibitem{33}
L.~Randall and R.~Sundrum,
Phys. Rev. Lett. {\bf 83}, 4690 (1999).
\bibitem{34}
J.~M.~Obrecht, R.~J.~Wild, M.~Antezza, L.~P.~Pitaevskii,
S.~Stringari, and E.~A.~Cornell,
Phys. Rev. Lett. {\bf 98}, 063201 (2007).
\bibitem {35}
G.~L.~Klimchitskaya and V.\ M.\ Mostepanenko,
J. Phys. A: Math. Theor. {\bf 41}, 312002 (2008).
\bibitem{36}
C.-C.~Chang, A.~A.~Banishev, R.~Castillo-Garza,
G.~L.~Klimchitskaya, V.\ M.\ Mostepanenko, and U.\ Mohideen,
Phys. Rev. B {\bf 85}, 165443 (2012).
\bibitem{37}
A.~A.~Banishev, C.-C.~Chang, R.~Castillo-Garza,
G.~L.~Klimchitskaya, V.\ M.\ Mostepanenko, and U.\ Mohideen,
Int. J. Mod. Phys. A {\bf 27}, 1260001 (2012).
\bibitem{38}
A.~A.~Banishev, C.-C.~Chang,
G.~L.~Klimchitskaya, V.\ M.\ Mostepanenko, and U.\ Mohideen,
Phys. Rev. B {\bf 85}, 195422 (2012).
\bibitem{39}
A.~A.~Banishev,
G.~L.~Klimchitskaya, V.\ M.\ Mostepanenko, and U.\ Mohideen,
Phys. Rev. Lett. {\bf 110}, 137401 (2013).
\bibitem{40}
A.~A.~Banishev,
G.~L.~Klimchitskaya, V.\ M.\ Mostepanenko, and U.\ Mohideen,
Phys. Rev. B {\bf 88}, 155410 (2013).
\bibitem{41}
R.~S.~Decca, D.~L\'opez, E.~Fischbach, G.~L.~Klimchitskaya,
 D.~E.~Krause, and V.~M.~Mostepanenko,
Phys. Rev. D {\bf 75}, 077101 (2007).
\bibitem{42}
R.~S.~Decca, D.~L\'opez, E.~Fischbach, G.~L.~Klimchitskaya,
 D.~E.~Krause, and V.~M.~Mostepanenko,
Eur. Phys. J. C {\bf 51}, 963 (2007).
\bibitem{43}
F.~Chen, U.\ Mohideen,
G.~L.~Klimchitskaya, and  V.\ M.\ Mostepanenko,
Phys. Rev. Lett.  {\bf 88}, 101801 (2002).
\bibitem{44}
F.~Chen, U.\ Mohideen,
G.~L.~Klimchitskaya, and  V.\ M.\ Mostepanenko,
Phys. Rev. A {\bf 66}, 032113 (2002).
\bibitem{45}
H.~B.~Chan, Y.~Bao, J.\ Zou, R.\ A.\ Cirelli, F.\ Klemens,
W.\ M.\ Mansfield, and C.\ S.\ Pai,
Phys. Rev. Lett. {\bf 101}, 030401 (2008).
\bibitem{46}
Y.~Bao, R.~Gu\'{e}rout, J.~Lussange, A.\ Lambrecht, R.\ A.\ Cirelli, F.\ Klemens,
W.\ M.\ Mansfield, C.\ S.\ Pai,
and H.~B.~Chan,
{Phys. Rev. Lett.} {\bf 105}, 250402 (2010).
\bibitem{47}
H.-C.\ Chiu,  G.~L.~Klimchitskaya, V.\ N.\ Marachevsky,
V.\ M.\ Mos\-te\-pa\-nen\-ko, and U.~Mohideen,
Phys. Rev. B {\bf 80}, 121402(R) (2009).
\bibitem{48}
H.-C.\ Chiu,  G.~L.~Klimchitskaya, V.\ N.\ Marachevsky,
V.\ M.\ Mos\-te\-pa\-nen\-ko, and U.~Mohideen,
Phys. Rev. B {\bf 81}, 115417 (2010).
\bibitem{49}
A.~A.~Banishev, J.~Wagner, T.~Emig,
R.~Zandi, and U.\ Mohideen,
Phys. Rev. Lett. {\bf 110}, 250403 (2013).
\bibitem{50}
A.~A.~Banishev, J.~Wagner, T.~Emig,
R.~Zandi, and U.\ Mohideen,
Phys. Rev. B {\bf 89}, 235436 (2014).
\bibitem{51}
G.~Vasilakis, J.~M.~Brown, T.\ R.\ Kornack, and
M.\ V.\ Romalis,
Phys. Rev. Lett. {\bf 103}, 261801 (2009).
\bibitem{52}
 E.~G.~Adelberger,
B.\ R.\ Heckel, S.\ Hoedl, C.\ D.\ Hoyle,
D.~J.~Kapner, and A.\ Upadhye,
Phys. Rev. Lett. {\bf 98}, 131104 (2007).
\bibitem{53}
D.~J.~Kapner, T.~S.~Cook, E.~G.~Adelberger,
J.\ H.\ Gundlach, B.\ R.\ Heckel, C.\ D.\ Hoyle,
and H.\ E.\ Swanson,
Phys. Rev. Lett. {\bf 98}, 021101 (2007).
\bibitem{54}
R.~Spero, J.~K.~Hoskins, R.~Newman, J.\ Pellam, and
J.~Schultz,
Phys. Rev. Lett. {\bf 44}, 1645 (1980).
\bibitem{55}
J.~K.~Hoskins, R.~D.~Newman, R.~Spero, and J.\ Schultz,
Phys. Rev. D {\bf 32}, 3084 (1985).
\bibitem{56}
G.~L.~Smith, C.~D.~Hoyle, J.~H.~Gundlach, E.~G.~Adelberger,
B.~R.~Heckel,  and H.~E.~Swanson,
Phys. Rev. D {\bf 61}, 022001 (1999).
\bibitem{57}
G.~G.~Raffelt,
J. Phys. A: Math. Theor. {\bf 40}, 6607 (2007).
\bibitem{58}
R.~S.~Decca, D.~L\'opez, E.~Fischbach,
 D.~E.~Krause, and C.~R.~Jamell,
Phys. Rev. Lett. {\bf 94}, 240401 (2005).
\bibitem{59}
M.~Masuda and M.~Sasaki,
{Phys. Rev. Lett.} {\bf 102}, 171101  (2009).
\bibitem{60}
S.~J.~Smullin, A.~A.~Geraci, D.~M.~Weld, J.~Chiaverini,
S.~Holmes, and A.~Kapitulnik,
Phys. Rev. D {\bf 72}, 122001 (2005).
\bibitem{61}
 A.~A.~Geraci, S.~J.~Smullin, D.~M.~Weld, J.~Chiaverini,
and A.~Kapitulnik,
{Phys. Rev. D} {\bf 78}, 022002 (2008).
\bibitem{62}
J.~H.~Gundlach, G.~L.~Smith, E.~G.~Adelberger,
B.~R.~Heckel,  and H.~E.~Swanson,
Phys. Rev. Lett. {\bf 78}, 2523 (1997).
\end{thebibliography}
\end{document}